# Dissipative particle dynamics simulations of weak polyelectrolyte adsorption on charged and neutral surfaces as a function of the degree of ionization

F. Alarcón[1,2], E. Pérez[1,3], and A. Gama Goicochea[†1,4]

[1]Centro de Investigación en Polímeros (Grupo COMEX) Marcos Achar Lobatón No. 2, Tepexpan, 55885 Acolman, Estado de México, Mexico

[2]Departament de Fisica Fonamental, Facultat de Fisica, Universitat de Barcelona, Marti i Franques, 1 08028 Barcelona, Spain

[3]Instituto de Física, Universidad Autónoma de San Luis Potosí, Álvaro Obregón 64, 78000 San Luis Potosí, Mexico

[4]Departamento de Ciencias Naturales, DCNI, Universidad Autónoma Metropolitana, Unidad Cuajimalpa, Av. Pedro Antonio de los Santos 84, México, D. F. 11850, Mexico

## Abstract

The influence of the chain degree of ionization on the adsorption of weak polyelectrolytes on neutral and on oppositely and likely charged surfaces is investigated for the first time, by means of Monte Carlo simulations with the mesoscopic interaction model known as dissipative particle dynamics. The electrostatic interactions are calculated using the three-dimensional Ewald sum method, with an appropriate modification for confined systems. Effective wall forces confine the linear polyelectrolytes, and electric charges on the surfaces are included. The solvent is included explicitly also and it is modeled as an athermal solvent for the polyelectrolytes. The number of solvent particles is allowed to fluctuate. The results show that the polyelectrolytes adsorb both onto neutral and charged surfaces, with the adsorption regulated by the chain degree of ionization, being larger at lower ionization degrees, where polyelectrolytes are less charged. Furthermore, polyelectrolyte adsorption is strongly modulated by the counterions screening of surface charge. These findings are supported with predictions of adsorption isotherms with varying ionization degree. We obtain also the surface force mediated by adsorbed polyelectrolytes, which is calculated for the first time as a function of ionization degree. The adsorption and surface force isotherms obtained for weak polyelectrolytes are found to reproduce main trends in experiments, whenever those results are available, and provide additional insight into the role played by the competitive adsorption of the counterions and polyelectrolytes on the surfaces.

---

[†] Corresponding author. Electronic mail: agama@correo.cua.uam.mx

I. INTRODUCTION

The adsorption of polyelectrolytes at surfaces is the basis of many water-based manufactured products and mineral recovery processes; also the behavior of many biological polyelectrolytes (such as proteins, or DNA) is influenced by the presence of charged membranes. The interest for polyelectrolyte adsorption on colloidal particles is especially relevant in the paint and coatings industry, where pigments have to be well dispersed to obtain optimal results in properties like gloss, opacity, and color distribution. These pigments are usually stabilized by polyelectrolytes, and the effectiveness of their adsorption is of capital importance when optimizing those properties [1]. From the point of view of fundamental research, several topics still remain poorly understood, such as the influence surface charge distribution on the adsorption, the competition among the counterions and polyelectrolytes for the adsorption sites, the dependence of these factors on the surface force, etc. (see, for example reference [2]). Many stimuli-responsive polymer brushes are polyelectrolytes [3]; hence the adsorption of polyelectrolytes at surfaces continues to be a subject of much experimental, theoretical and computer simulation research.

Polyelectrolytes are classified by their ability to be ionized in water as strong or weak polyelectrolytes. Strong polyelectrolytes are easily ionized, and electrostatic charge is not too sensitive to pH in solution. By contrast, weak polyelectrolytes are very sensitive to pH changes because ionic equilibrium determines the ionization of their anionic/cationic groups [4]. The pH value essentially controls the degree of ionization ($\alpha$) of weak polyelectrolytes, which is directly related to the dissociation constant of the anionic group and the polyelectrolyte conformation [5]. In addition, the properties of colloidal metallic particles in aqueous suspensions are usually pH dependent also, because metallic oxides are present on their surface. Weak polyelectrolyte adsorption onto pigment surfaces (say, $TiO_2$), to name an example, is thus pH dependent, and colloidal stability is frequently associated with a limiting pH value in these dispersions [6].

The adsorption of weak polyelectrolytes onto colloidal pigments, as a function of pH and ionic strength, has been investigated experimentally for different metals whose surfaces are usually oxidized, such as polyacrylic acid, polyacrilamide, and modified polyacrilamides onto titania [7]; polyacrylic acid onto alumina [8]; and carboxymethyl cellulose, polyacrylamide, and modified polyacrylamides onto talc [9]. Pigments are usually characterized by their isoelectric point, which is the point of zero charge in an titration experiment: at low pH (acid solutions) the oxide surface is positively charged, while at high pH (basic solutions) it is negatively charged. These works have shown that at low pH values, the adsorption of anionic polyelectrolytes on positively charged surfaces is high, while at higher pH values the adsorption on negatively charged surfaces decreases. Polyelectrolyte adsorption onto surfaces with charges of sign opposite to those on the polyelectrolyte at low pH is expected, but polyelectrolyte adsorption on surfaces with the same charge at high pH is not, simply because of Coulomb repulsion, at least if no other mechanism is at work. However, such same-charge adsorption has been observed and it has been explained by the presence of counterions that decrease the electrostatic interaction between the polyelectrolyte and pigment [10], or even by the possible presence of H-bonds between polyelectrolyte groups and pigment surfaces [7, 11]. Moreover, the role of ions in the adsorption of polyelectrolytes has been also recognized. These ions may produce a salt-exclusion effect that limits the polymer adsorption at high surface charge [10], and define low or high-salt regimes, where the electrostatic interactions are long or short ranged, and satisfy different scaling laws [12].

These trends have been observed in the adsorption of polyacrylic acid on strong, positively charged polystyrene latex by quaternary ammonium groups, whose ionization is independent of pH [11]. In this system, upon pH changes, the degree of ionization of the polyelectrolyte is varied, but the sign of the charge is conserved. In such case, the polyelectrolyte adsorption also increases as the pH decreases, arriving at a high adsorption when the pH of the solution is close to the $pK_0$ of the polyelectrolyte, which can be explained using an extension of the self-consistent-field theory of Scheutjens and Flee [11]. There are now a number of theoretical works that have advanced significantly the field of polyelectrolyte adsorption. Dobrynin and co workers [13] developed a scaling theory for the adsorption of polyelectrolytes at oppositely charged surfaces, and found that the adsorbed polyelectrolyte layer thickness was the results of the balance between electrostatic attraction and chain entropy, for weakly charged surfaces. However, they did not address the issue of the polyelectrolyte adsorption when the pH is controlled, nor did they study the effect of the interaction of the polymer backbone and the surface. On the computer modeling side, Monte Carlo (MC) simulations have been used to study the conformational changes of one polyelectrolyte in the presence of one oppositely charged particle and to predict adsorption isotherms of ionic surfactants on neutral particles as a function of the surfactant-particle *non-electrostatic* interaction, finding similar trends to those described above [14]. Using constant density MC simulations, Messina [15] found that the effect of image forces can reduce polyelectrolyte adsorption and consequently, may inhibit charge inversion on the surfaces due to the polyelectrolytes. The influence of pH, or that of the confinement (at fixed chemical potential) on polyelectrolyte adsorption was not discussed in that work. On the other hand, Velichko and Olvera de la Cruz [16] found that polyelectrolytes can form patterns on charged surfaces, while Chang and Yethiraj [17] used molecular dynamics simulations to model polyelectrolyte association, and found that the inclusion of the solvent explicitly was important to allow or inhibit the formation of certain structures. However, we are not aware of any simulation work on the prediction of adsorption and surface force isotherms of polyelectrolytes as a function of pH or ionization degree on charged and neutral surfaces.

In the present work, we study the adsorption of weak positively charged polyelectrolytes onto neutral and negatively charged surfaces, and of weak negatively charged polyelectrolytes onto a negatively charged surface, as a function of the degree of ionization, by means of MC simulations in the semi Grand Canonical (GC) ensemble (hereafter referred to for simplicity as GCMC), namely, at fixed solvent's chemical potential ($\mu$), volume ($V$), and temperature ($T$). The surface force that the colloidal particles experience is also calculated, to our knowledge, for the first time, as a function of the degree of ionization. The interaction model used is the coarse-grained, mesoscopic method known as dissipative particle dynamics (DPD) [18], which is complemented with the explicit inclusion of electrostatic interactions by means of the Ewald sums, appropriately adapted for confined systems [19]. The short-range interactions among the fluid particles and the colloids are modeled by exact, effective DPD walls [20]. Additionally, a coarse-grained model for the charges distributed on the colloidal particle surface is proposed. The solvent and counterions are included explicitly.

This work is organized as follows. In Section II we show very briefly the DPD interaction model; Section III is devoted to a review of the GCMC-DPD hybrid algorithm for confined fluids. In Section IV one can find the details about how the electrostatics is included in a confined DPD fluid, with surface charges as detailed in Section V. The computational details of our simulations as well as the description of all systems studied are to be found in Section VI, while the results and discussion are presented in Section VII. Finally, conclusions are drawn in Section VIII.

## II. DPD FORCE MODEL

In DPD, the fluid is coarse-grained into soft, momentum carrying beads, with their dynamics governed by the following stochastic differential equation [18]:

$$\dot{v}_i = \sum_{j \neq i} \left( F_{ij}^C + F_{ij}^D + F_{ij}^R \right) \tag{1}$$

$$\dot{r}_i = v_i, \tag{2}$$

where all masses are set equal to 1. The pairwise, additive forces in equation (1), are identified as the conservative $F_{ij}^C$, dissipative $F_{ij}^D$, and random $F_{ij}^R$ contributions, and are given by:

$$\vec{F}_{ij}^C = a_{ij} \left( 1 - r_{ij}/R_c \right) \hat{e}_{ij} \tag{3}$$

$$\vec{F}_{ij}^D = -\gamma \left( 1 - r_{ij}/R_c \right)^2 [\hat{e}_{ij} \cdot \vec{v}_{ij}] \hat{e}_{ij} \tag{4}$$

$$\vec{F}_{ij}^R = \sigma \left(1 - {r_{ij}}/{R_c}\right) \hat{e}_{ij} \xi_{ij} \qquad (5)$$

where $r_{ij} = r_i - r_j$ is the relative position vector, $\hat{e}_{ij}$ is the unit vector in the direction of $r_{ij}$, and $v_{ij} = v_i - v_j$ is the relative velocity, with $r_i, v_i$ the position and velocity of particle *i*, respectively. The random variable *ξ*$_{ij}$ is generated between 0 and 1 with Gaussian distribution of unit variance; *a*$_{ij}$, *γ* and *σ* are the strength of the conservative, dissipative and random forces, respectively; $R_c$ is a cut off distance. All these forces are zero for $r_{ij} > R_c$. All beads are of the same size, with radius $R_c$, which is set equal to 1. The constants in equations (4) and (5) must satisfy:

$$\frac{\sigma^2}{2\gamma} = k_B T \qquad (6)$$

as a consequence of the fluctuation – dissipation theorem [17], where *k*$_B$ is Boltzmann's constant. The short-range nature of these forces, and their linearly decaying spatial dependence, allow the use of relatively large time steps when integrating the equation of motion. As the DPD particles are representations of groups of atoms or molecules, the DPD method becomes an attractive alternative to study systems at the mesoscopic level. Note also that the forces in equations (3)-(5) obey Newton's third law, which means momentum is conserved locally, and globally, which in turn preserves any hydrodynamic modes present in the fluid. Another appealing feature of DPD is the natural thermostat that arises from the compensation between the dissipative and random forces, as stated in equation (6). This interaction model has been used successfully to predict equilibrium properties of polymer melts [22], surfactants in solution [23], and colloidal stability [24], to name a few. For further reading, see reference [25].

### III. GCMC-DPD FOR CONFINED SYSTEMS

We use the GC (or *μVT*) ensemble to ensure the fluid modeled is in chemical, mechanical and thermal equilibrium. The algorithm, which we call GCMC-DPD [24] is a hybrid between MC and dynamics, because for each MC cycle one performs 10 DPD steps, integrating equations (1) and (2) to create new positions and velocities for the fluid particles. After the tenth step of the dynamics, the *total* energy of the system is compared with the initial energy. Then, the new state is accepted or rejected using the Metropolis algorithm [26]. When this process has been completed, the exchange of monomeric solvent beads with the bulk is attempted, inserting or deleting beads to keep the chemical potential fixed, with probability (here, *min*[*a*,*b*] indicates that the minimum between *a* and *b* is to be chosen):

$$P_{insertion} = min\left[1, \frac{\langle Z(z)\rangle V}{N+1} exp\left(-\frac{\Delta U^{test}}{k_B T}\right)\right] \quad (7)$$

$$P_{deletion} = min\left[1, \frac{N}{\langle Z(z)\rangle V} exp\left(-\frac{\Delta U^{test}}{k_B T}\right)\right] \quad (8)$$

where $\Delta U^{test}$ is the conservative interaction energy difference between the added or removed bead, and the $N$ or $N-1$ remaining beads, including the conservative interaction with the surfaces. $\langle Z(z)\rangle$ is the so-called activity, defined for inhomogeneous systems as [27]:

$$\langle Z(z)\rangle = \frac{\langle \rho(z)\rangle_{NVT}}{\langle \exp\left(-\Delta U^{test}/k_B T\right)\rangle_{NVT}} = e^{\langle \mu(z)\rangle/k_B T} \quad (9)$$

where <$\mu(z)$> is the chemical potential for an inhomogeneous system in the *z*-direction, and the angular brackets represent averages in the *NVT* (canonical) ensemble. The total average, *z*-dependent density is $\rho(z)$. To keep the chemical potential of the solvent constant, the addition or removal of solvent beads is performed a number of times approximately equal to the average number of DPD beads. Once this interchange is completed, a new MC cycle begins until equilibrium has been attained, and stable acceptance rates have been reached. Afterwards, averages of the equilibrium thermodynamic properties are obtained and analyzed. The constant chemical potential is chosen so that the averaged fluid density was <$\rho^*$> ~3 [28]. This ensemble is sometimes called semi Grand Canonical. Our model is based on the assumption that when equilibrium is reached, the chemical potential of the ions and polymers are equal to the solvent's chemical potential. This is an appropriate model because the polymer adsorption process has been completed when the system reaches equilibrium and the ions remain close to them due to Coulomb attraction. Full details of this hybrid, GCMC-DPD algorithm can be found in reference [24].

An exact expression for the effective force exerted by a uniform and homogeneous surface made up of DPD beads in contact with a DPD fluid is used here, to model colloidal particles on which polyelectrolytes can be adsorbed. It was derived [20] by calculating the interaction between an infinite surface, made up of a uniform distribution of DPD beads, and free fluid DPD particles in the semi-infinite space in the *z*-direction. The effective force that the DPD wall exerts on the fluid particle (*i*) is exactly given by:

$$\boldsymbol{F}_i^w(z) = a_w \left[ 1 - 6\left(\frac{z}{R_c}\right)^2 + 8\left(\frac{z}{R_c}\right)^3 - 3\left(\frac{z}{R_c}\right)^4 \right]. \tag{10}$$

In the equation above, $a_w = \frac{1}{12} a_{ij} \pi \rho_w R_c^3$, with $a_{ij}$ being the strength of the DPD conservative interaction between a particle in the fluid and a surface particle (see equation 3), $\rho_w$ is the density of the wall, and the force vanishes for distances $z > R_c$. Additionally, polymers are modeled as linear chains formed by DPD beads joined by freely rotating, harmonic springs with spring constant $K_0 = 100.0$ and equilibrium position $r_0 = 0.7$ [29], as shown in the following equation.

$$\boldsymbol{F}_{ij}^{spring} = -K_0 (r_{ij} - r_0) \hat{\boldsymbol{e}}_{ij}. \tag{11}$$

The interaction potential from which the wall force in equation (10) above derives, and the harmonic spring energy for polymers, must be included when calculating energy differences between MC configurations, be it under *NVT* or *µVT* conditions. The electrostatic contribution to the energy must also be included, which is introduced as follows.

## IV. ELECTROSTATICS FOR CONFINED DPD FLUIDS

Electrostatic interactions have been introduced into the DPD model using two different approaches. One consists of distributing charges on a lattice and then solving the Poisson – Boltzmann equation [30]; the other [31], which we use here, is by means of the standard Ewald method [32], adapted for DPD using charge distributions with spherical symmetry and a Slater-type, radial exponential decay with decay length $\lambda$, instead of point charges. This charge distribution is adopted because the soft-core interaction of the DPD beads allows the collapse of point charges and the subsequent formation of artificial ionic pairs. Following references [33] and [34], the charge distribution used here is given by:

$$\rho(r) = \frac{q}{\pi\lambda^3} e^{-\frac{2r}{\lambda}}. \tag{12}$$

The dimensionless force between two charge distributions is given by two contributions [34], one in real space ($\boldsymbol{F}_{ij}^{E,R}$), plus one in Fourier space ($\boldsymbol{F}_{i}^{E,K}$):

$$\boldsymbol{F}_{ij}^{E,R} = \frac{\Lambda}{4\pi} Z_i Z_j \left[ \frac{2\pi}{V} exp(-\alpha^2 r_{ij}^2) + erfc(\alpha r_{ij}) \right] \\ \times \{1 - exp(-\beta r_{ij})[1 + 2\beta r_{ij}(1 + \beta r_{ij})]\} \frac{\boldsymbol{r}_{ij}}{r_{ij}^3}, \tag{13}$$

$$\boldsymbol{F}_{i}^{E,K} = \frac{\Lambda}{4\pi} Z_i \left\{ \frac{2\pi}{V} \sum_{\boldsymbol{k}\neq 0}^{\infty} Q(k) \boldsymbol{k} \times Im[exp(-i\boldsymbol{k}\cdot\boldsymbol{r}_i) S(\boldsymbol{k})] \right\}, \tag{14}$$

where $\Lambda = e^2/(k_B T \varepsilon_0 \varepsilon_r R_c)$, with $e$ being the electron charge, $\varepsilon_0$ is the permittivity of vacuum, $\varepsilon_r = 78.3$ is the water relative permittivity at room temperature; $\beta = R_c / \lambda$, and $Z_i$ is the valence of the charge distribution; $\boldsymbol{k} \equiv 2\pi (k_x/L_x, k_y/L_y, k_z/L_z)$ is the reciprocal vector of magnitude $k$, such that $k_x$, $k_y$, and $k_z$ are integers. $Q(k) = {exp\left(-\frac{k^2}{4\alpha^2}\right)}/{k^2}$, and $S(\boldsymbol{k})$ is the so-called structure factor [32]. The term $erfc(x)$ is the complementary error function; $\alpha$ is the parameter that modulates the contribution of the sum in real space (not to be confused with the ionization degree). $V=L_xL_yL_z$ is the volume of the simulation box. Here, *Im* denotes the imaginary part of the complex number.

The forces given by equations (13) and (14) are appropriate for the calculation of the long-range electrostatic interactions of a homogeneous, non confined system (which we shall call Ewald-3D), but as they stand, they cannot be used for confined systems, which are the subject of this work. The reason for this relies on the fact that the Fourier transformation involved in equation (14) cannot be performed straightforwardly if the system lacks 3D periodicity. However, Yeh and Berkowitz [35] have proposed a computationally inexpensive modification so that they can be applied to inhomogeneous systems (called hereafter Ewald-3D_C). In this method, a *z*-component force is added to each charged bead:

$$\boldsymbol{F}_{i,z} = -\frac{\Lambda q_i}{V} M_Z, \tag{15}$$

where $M_Z$ is the net dipole moment of the simulation cell, which is given by

$$M_Z = \sum_{i=1}^{N} q_i z_i \ ; \tag{16}$$

such dipole moment must be removed out of the simulation cell for each charge $q_i$.

The importance of including the force shown in equation (15) can be ascertained from Figure 1, where the *z*-component of the force acting on a charged DPD bead located at (0, 0, *z*) by another, oppositely charged bead fixed at (0, 0, 0), is shown. Because of the periodicity in the *xy*-plane, the charges form charge sheets, whose force on one another should be *z*-independent (see equation (15)). The forces were calculated using both Ewald-3D and Ewald-3D_C methods, and are shown in Figure 1 for two simulation box sizes, for both methods.

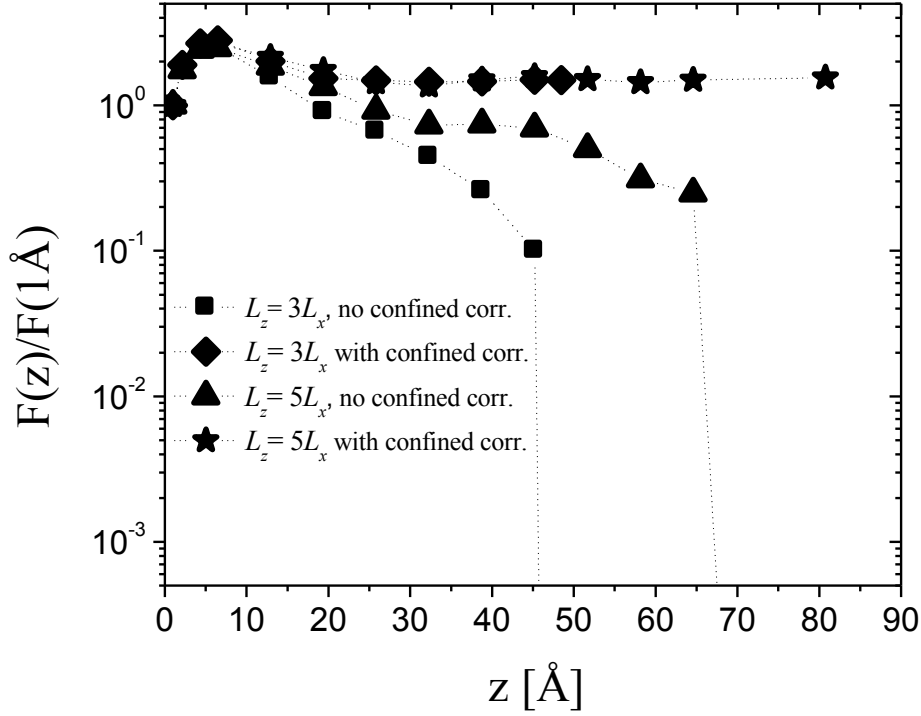

**Figure 1.** Comparison of the *z* component of the electrostatic force acting on 2 DPD beads with opposite charge distributions, with one located at (0, 0, *z*) and the other at (0, 0, 0) in a system periodic in the *xy*-plane, calculated using 2 different Ewald methods, and for 2 different box sizes. The sides of the simulation box in the *xy*-plane are equal, i. e., $L_x=L_y$, with $L_z$ given as indicated in the legend. The position z=$z^*R_c$, with $R_c$ =6.46 Å, see text for details. The solid squares (when $L_z$ = 3 $L_x$) and solid triangles (for $L_z$ =5 $L_x$) correspond to the use of the Ewald sums for systems with 3-dimensional symmetry ("no confined corr."), and lead to incorrect values of the force as $z\rightarrow\infty$. Solid diamonds (for $L_z$ = 3 $L_x$) and solid stars (for $L_z$ = 5 $L_x$) represent the force calculated using the 3-dimensional Ewald sums with the addition of equation (15) to account for the confinement ("with confined corr."). The lines are only guides to the eye. See text for details.

One can easily notice that when the full, Ewald-3D sums are used without the proper correction to account for the confinement, (solid squares and triangles, in Figure 1) an incorrect limiting behavior for the force as the distance *z* grows is obtained, since the force tends to 0. By contrast, when the correction for confinement (equation (15)) is included, as shown for the (small) solid diamonds and (larger) solid stars simulation boxes in Figure 1, the force between the charge sheets has the correct limiting behavior as the distance separating the sheets, z, grows. Another feature, which is intrinsic to the charge distribution model used here, is the "hump" that appears in the forces shown in Figure 1 at distances less than 20 Å. It arises from the spatial dependence of the electrostatic force between 2 charge distributions of the type shown in equation (12), see reference [31]. It has been shown [35] that the Ewald-3D_C method, which we adopt here, leads to results that are almost identical to the exact 2-dimensional, and computationally much more expensive, version of the Ewald sums, [36]. We have shown, for the first time, that the same result is obtained for interactions between charge distributions, rather than point charges. Only the 3-dimensional method with correction for confined systems (Ewald-3D_C) shall be used in the rest of this work.

## V. SURFACE CHARGE DISTRIBUTION MODEL

Many colloidal particles as well as biological membranes have surface charges that need to be taken into account explicitly, because this is an additional variable that can be used experimentally to tune up properties. In keeping with the mesoscopic, DPD approach we introduce here a surface charge distribution, taking as a starting point the effective electrostatic force arising from point charges on a uniform, planar $TiO_2$ surface interacting with point charges in the fluid [37]. The effective electrostatic force between the $TiO_2$ wall and the fluid's particle $i$ is given by

$$F_i^{EW}(z) = \frac{q_i}{\epsilon_0} \frac{\rho_{OX} a_z^2}{z} \qquad (17)$$

where $q_i$ is the electric charge of fluid particle $i$, $\rho_{OX}$ is the charge density of the oxygen atoms on the wall, and $a_z$ is the z-component of the $\vec{a}$ vector ($\vec{a}$ is the bond vector from the Ti atom to each O atom in the $TiO_2$ molecule). Equation (17) is the force for point particles, not distributions of charge. Now, if one models the charge density of the oxygen atoms by a charge distribution of the Slater type, which is similar to equation (12), one then writes:

$$\rho_{OX}(z) = \frac{q_{OX}}{\pi \lambda^3} e^{-\frac{2z}{\lambda}}; \qquad (18)$$

where $q_{OX}$ is the charge of the oxygen atom. The electrostatic interaction potential between a fluid particle and the bead on the wall, in dimensionless units, is:

$$U_i^{EW}(z) = -\Lambda \kappa Z_i e^{-2\beta z} \ln(z), \quad (19)$$

therefore, for a charge distribution on the surface and a charged DPD fluid particle, the electrostatic force is

$$\boldsymbol{F}_i^{EW}(z) = \Lambda \kappa Z_i e^{-\frac{2z}{\lambda}} \left[\frac{1}{z} - 2\beta \ln(z)\right], \quad (20)$$

where $\Lambda$ and $\beta$ were defined before (see text following equation (14)) and

$$\kappa = \frac{a_z^2 Z_{ox}}{\pi \lambda^3}. \quad (21)$$

Here, $Z_i$ is the valence of the fluid particle $i$, $Z_{ox} = -1.15$ (valence of the two oxygen atoms in the TiO$_2$ molecule), and $a_z = 1.27$ Å $= 0.197\ R_c$.

The total force $\boldsymbol{F}_i$ acting on bead $i$ is the sum of the DPD forces (equations (3)-(5)), the force due to the DPD surface (equation (10)), the spring force between beads in a polyelectrolyte (equation (10)), the forces related to the electrostatic interactions (equations (13)-(15)) and the force that takes into account the charge distribution on the surface (equation (20)). Therefore, equation (1) can be rewritten as:

$$\begin{aligned}\dot{\boldsymbol{v}}_i = \boldsymbol{F}_i = \sum_{j \neq i} \boldsymbol{F}_{ij}^C + \boldsymbol{F}_{ij}^D + \boldsymbol{F}_{ij}^R + \boldsymbol{F}_i^W(z) + \boldsymbol{F}_{ij}^{spring} + \boldsymbol{F}_{ij}^{E,R} + \boldsymbol{F}_i^{E,K} + \boldsymbol{F}_{i,Z} \\ + \boldsymbol{F}_i^{EW}.\end{aligned} \quad (22)$$

This is the equation that is integrated to obtain the particles' positions and velocities at every time step in the dynamics part of the GCMC-DPD algorithm. Additionally, every time the Metropolis algorithm is applied (in the *NVT* and *μVT* ensembles), the full conservative energy is used, that is, the one that includes the DPD potential energy, the harmonic potential, the electrostatic potential energy with its confinement correction, and the surface charge potential.

## VI. SYSTEMS STUDIED AND SIMULATION DETAILS

We carried out 3 types of simulation studies: adsorption of cationic polyelectrolytes on neutral and negatively charged surfaces, and adsorption of anionic polyelectrolytes on negatively charged surfaces, as a function of the degree of ionization (which is related to the pH value). Full surface force isotherms of the cationic polyelectrolytes confined by two negatively charged surfaces are also obtained as a function of ionization degree. Dimensionless units are used throughout this work, except where indicated otherwise. When calculating adsorption isotherms, the volume of the simulation box was fixed with $L_x = 5$, $L_y = 5$, $L_z = 10$, and periodic boundary conditions were used, except in the $z$-direction since this is direction of the confinement.

For the calculation of the surface force isotherms we fixed the polyelectrolyte saturation concentration obtained from the adsorption isotherms, and performed simulations at varying separation between the surfaces, in the $z$-direction (keeping the transversal area equal to that used in the calculation of adsorption isotherms). Under these conditions, all polyelectrolyte molecules are adsorbed, and the counter ions remain close to them at equilibrium. Therefore, only the solvent's chemical potential was set, and calculated also as a check on self-consistency, using the Widom insertion method [32]. The chemical potential was fixed at $\mu^* = 37.7$, which gives a total average density of $<\rho^*> \approx 3$. The simulation results were obtained after at least 100 blocks, of $10^4$ MC configurations each, with the first 30 blocks used to equilibrate the system. In each given block made up of $10^4$ MC configurations, the percentage of successful MC moves was around $(35 \pm 2)$ %. The time step was $\delta t^* = 0.03$ for the DPD part, where the DPD beads were moved using the Velocity Verlet algorithm, DPD – VV [38]. The parameters that define the dissipative and random forces intensities are $\gamma = 4.5$ y $\sigma = 3.0$, so that $k_B T^* = 1$. The conservative force intensities were chosen as $a_{ij} = 78.0$, when $i = j$, and $a_{ij} = 79.3$, when $i \neq j$, where $i$ and $j$ represent types of fluid molecules (solvent, polymer or counterion). With this choice of parameters, the polyelectrolytes are in an athermal solvent. Additionally, $a_w = 120.0$ for the solvent – wall non-electrostatic interaction, and $a_w = 100.0$ for all the other molecular

species – wall interactions. If the parameter $a_w$ were to be chosen equal for solvent particles, as well as for polyelectrolytes and ions, all of these particles would be equally likely to adsorb. Our choices of wall – particle interactions necessarily lead to adsorption of polyelectrolytes and ions on the surfaces, since one does not need attractive interactions for adsorption to occur, only less repulsive interactions with respect to the interaction of the wall with the particles that do not adsorb, in our case, the solvent [24]. The solvent and the counterions are modeled as simple monomeric particles. All polyelectrolytes are linear molecules made up of 7 DPD beads freely joined by harmonic springs. Only up to four of those seven beads were allowed to have electrical charge (chosen in alternating order) and the remaining ones were maintained neutral in the simulations; this is how the ionization degree was fixed. The center of mass of the charged distribution was fixed at the center of mass of the polyelectrolyte bead. These conditions limit the electrical charge as occurs for weak polyelectrolytes. The parameters for the Ewald sums were $\alpha = 0.11$ Å$^{-1}$ and the maximum vector $\boldsymbol{k}^{\max} = (6,6,6)$. The values of $\beta = 0.929$, $\lambda = 6.95$ Å and $\Lambda = 13.87$ were used, as in reference [31]. The charged surfaces were neutralized by the cationic polyelectrolytes, or by the positive counterions when the polyelectrolytes were anionic. Lengths were normalized_with $R_c = 6.46$ Å, which is the value that corresponds to the coarse graining degree of 3 water molecules in one DPD bead [30]. This length is also the maximum range of the interparticle forces (see equations (3)-(5)) and the non-electrostatic wall force (see equation (10)).

## VII. RESULTS AND DISCUSSION

The purpose of these simulations is the prediction of adsorption isotherms of weak polyelectrolytes onto neutral surfaces and onto oppositely and a likely charged surfaces upon change in the degree of ionization. We want also to obtain surface force isotherms of polyelectrolytes adsorbed on oppositely charged surfaces as a function of the ionization degree of the solution. The influence of the interparticle conservative interaction ($a_{ij}$, see equation (3)) and that of the model used for the effective confining walls (see equation (10)) on the characteristics of the adsorption and disjoining pressure isotherms (the latter related to surface force isotherms) have been studied in detail in a previous work [24].

The polyelectrolyte concentration ($\Gamma$) at the surface as a function of the polymer concentration in the bulk (not adsorbed) in equilibrium, was calculated from the density profile $\rho(z)$ through equation (23):

$$\Gamma = \int [\rho(z) - \rho_b] dz, \qquad (23)$$

where $\rho_b$ is the polyelectrolyte density in the bulk and the integral is performed over all the coordinate $z$ ($L_z = 10$ in all the cases). The polyelectrolyte bulk density, $\rho_b$, was obtained for each polyelectrolyte concentration simulated from the corresponding density profile obtained at the particular polyelectrolyte concentration, $\rho(z)$, taking $\rho_b$ as the average value reached when the perturbations of the walls on the polyelectrolyte profile were negligible [24], i. e., at the center of the simulation box, see Figure 2. Then, equation (23) is applied. This is also the procedure followed experimentally to obtain adsorption isotherms. These isotherms were fit to the Langmuir model, given by:

$$\Gamma = \frac{\Gamma_{max} K C_b}{1 + K C_b}, \qquad (24)$$

where $C_b$ is the concentration of polymer in the bulk (not adsorbed), $\Gamma_{max}$ is the maximum polyelectrolyte adsorption, and $K$ is the equilibrium constant of the Langmuir model [39]. In Figure 2 we show the density profiles obtained for one of the systems we study in this work, namely the adsorption of anionic polyelectrolytes on negatively charged surfaces, at 3 different degrees of ionization. Only the density profiles near the left surface are shown for simplicity, since they area symmetrical with respect to the right surface. The adsorbed polyelectrolytes form layers that appear as maxima in Figure 2, while the non adsorbed polyelectrolytes constitute a featureless bulk ($\rho_b$), represented by the constant region of the density profile (for $z^*$ larger than about 3 in Figure 2).

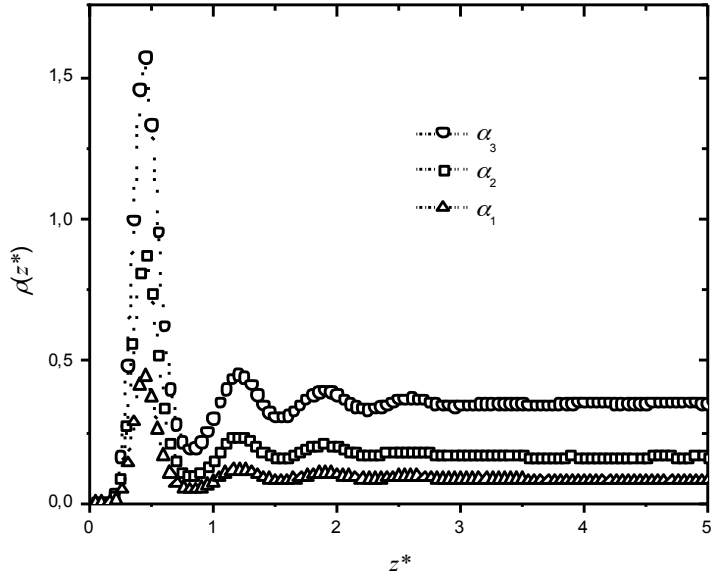

**Figure 2.** Segment density profiles of anionioc polyelectrolytes on negatively charged surfaces for $\alpha_1$ (Δ), $\alpha_2$ (□) and $\alpha_3$ (○). Due to symmetry, only the density profiles close to the left wall are shown, and the solvent and counterions density profiles are omitted, for clarity.

In these simulations, we have modeled positively charged polyelectrolytes onto neutral and negatively charged surfaces, and negatively charged polyelectrolytes onto a negatively charged surface. Clearly, the relation between the chain degree of ionization ($\alpha$) and the pH value is opposite for anionic and cationic polyelectrolytes: $\alpha$ decreases (increases) with pH for cationic (anionic) polyelectrolytes. To compare our predictions with experiments, we relate the degree of ionization of the polyelectrolytes with their pH using an extension of Henderson-Hasselbalch equation for polyelectrolytes [40]. In this approach, the pH of the solution is related to the chain degree of ionization ($\alpha$) by:

$$pH = pK_a \mp log_{10}\left(\frac{\alpha}{1-\alpha}\right) \qquad (25)$$

where the negative sign is used for cationic polyelectrolytes, and the positive sign for anionic ones, and $pK_a$ is the apparent ionic constant, with $pH = pK_a$ when $\alpha = 0.5$. Once the charge on each bead of the polyelectrolyte is chosen, the pH variation is evaluated using equation (25).

Equation (25) is commonly used to calculate the dissociation of weak anionic polyelectrolytes in solution [40]. In the present work, we use equation (25) to obtain the main functional dependence of the degree of ionization with the pH value for both cationic and anionic polyelectrolytes; but it must be stressed that the degree of ionization is equal to the degree of dissociation only for the anionic polyelectrolytes. For polyelectrolytes near a surface, equation (25) should be improved so that it takes into account at least two effects: the $pK_a$ dependence on both $\alpha$ and the polyelectrolyte molecular weight $N$ [40], and the variation of $\alpha$ near a charged surface [11]. For negative polyelectrolytes, the $pK_a$ value is shifted to high pH values as function of $\alpha$; scaling arguments give an approximation for this contribution, which is $\Delta pK = pK_a - pK_0 \sim \alpha^{1/3}(\ln N)^{2/3}$, with $pK_0$ the intrinsic $pK_a$ of a monomer [41]. The shift is not significant at low degree of ionization and small molecular weight, which are the conditions of the present simulations. Additionally, the variation of $\alpha$ for the polyelectrolyte near a charged surface has been calculated in reference [11]. This effect is more important for the polyelectrolyte segments adsorbed on the surface, and $\alpha$ is greater (smaller) for adsorbed polyelectrolyte segments than for polyelectrolytes in solution at low (high) pH values, indicating a prolongation in the titration curve that extends the dependence of the negative polyelectrolytes, adsorbed on the charged surface, at low and high pH values. Neutral surfaces only produce a $pK_a$ shift to high pH values. Therefore, in regard to the simulations presented here, these effects do not modify the main trends obtained from equation (25) when applied to polyelectrolytes adsorbed on a surface. Similar trends are expected for cationic polyelectrolytes but with the opposite pH dependence.

The results are then presented in terms of 3 pH values, namely $pH_1$, $pH_2$, and $pH_3$, which correspond to degrees of ionization of $\alpha_1 = 0.14$, $\alpha_2 = 0.29$, and $\alpha_3 = 0.57$, respectively. According to equation (25), as the degree of ionization ($\alpha$) increases, the pH decreases for cationic polyelectrolytes, and it increases for anionic polyelectrolytes. Thus, $pH_1 > pH_2 >$

pH$_3$ for cationic polyelectrolytes and pH$_1$ < pH$_2$ < pH$_3$ for anionic ones. Note that the $pK_a$ value is between pH$_2$ and pH$_3$ for both types of polyelectrolytes.

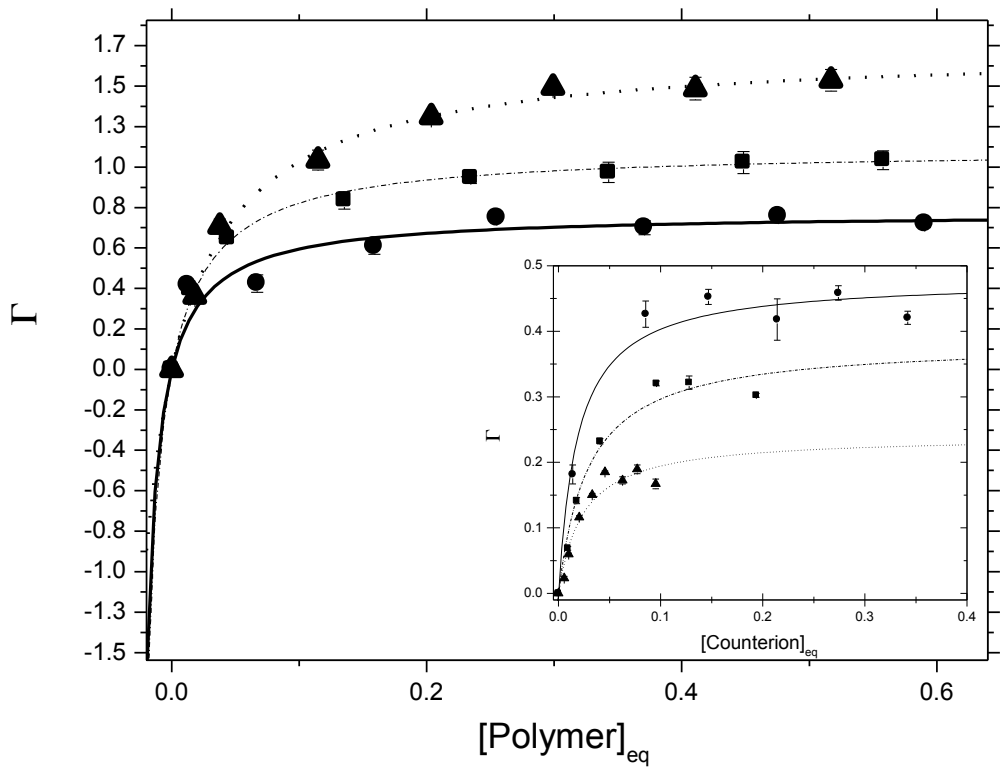

**Figure 3.** Adsorption isotherms of cationic polyelectrolytes on neutral surfaces for $\alpha_1$ (▲), $\alpha_2$ (■), and $\alpha_3$ (●). As described in the text, $\alpha_1 < \alpha_2 < \alpha_3$, which correspond to pH$_1$ > pH$_2$ > pH$_3$ in this case. The adsorption isotherms of the counterions are shown in the inset. The bars represent the statistical uncertainty. The continuous lines represent the data fitting to the Langmuir model, equation (24).

The adsorption isotherms of the cationic polyelectrolytes and negative counterions on a neutral surface are shown in Figure 3. The only interaction between the surfaces and the polymers is through $F_i^w(z)$, given by equation (10), but the adsorption is effectively influenced by the degree of ionization. The greater the ionization, the lower the polyelectrolyte adsorption, or equivalently: the cationic polyelectrolyte adsorption increases as the pH is increased (recall that pH$_1$ > pH$_2$ > pH$_3$, in this case). Indeed, it is not influenced by the fact that the $pK_a$ is between pH$_2$ and pH$_3$. As observed in Figure 3, the counterions adsorption follows a behavior just opposite to the polyelectrolyte adsorption; it increases with decreasing pH. Evidently, both phenomena are related. As the polyelectrolyte is dissociated the counterions are preferentially adsorbed onto the neutral surface ¡thereby preventing the polyelectrolyte adsorption. Good agreement is found between our simulations and Langmuir's model, as indicated in Figure 3 by the lines. The same trends are observed experimentally in the adsorption of polyethylenimine (PEI), a positively charged polyelectrolyte, on a neutral surface (graphite) at different pH values (3, 6, and 11) characterized with atomic force microscopy (AFM) [42]. At pH = 3, PEI is highly charged and shows a low adsorption, which then increases at pH = 6, and even more at pH = 11, thus adsorption is increased with pH. The agreement between these experimental results and our predictions in Figure 3 is therefore excellent.

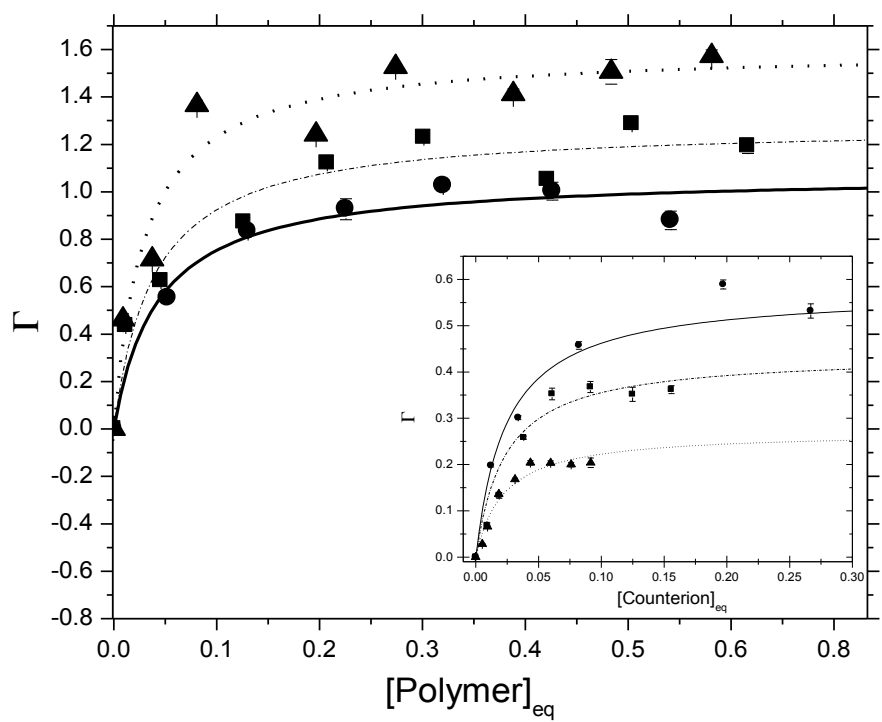

**Figure 4.** Adsorption isotherms of cationic polyelectrolytes on negatively charged surfaces for $\alpha_1$ (▲), $\alpha_2$ (■) and $\alpha_3$ (●). As in the previous figure, $\alpha_1 < \alpha_2 < \alpha_3$ with $pH_1 > pH_2 > pH_3$. The adsorption isotherms of the counterions are shown in the inset. The bars represent the statistical uncertainty. The continuous lines represent the data fitting to the Langmuir model, Equation (24).

The adsorption isotherms of cationic polyelectrolytes (and those of the negative counterions), when the surfaces are negatively charged, are shown in Figure 4. In this case, the electrostatic attraction between the wall and the polyelectrolytes is given by equation (20), and the repulsive DPD interaction between the walls and the fluid particles, including the polyelectrolytes, is given by equation (10). The competition between these forces gives rise to the isotherms in Figure 4, which follow the same trends with pH as those of the neutral wall, namely the cationic polyelectrolyte adsorption increases when the pH increases and the degree of ionization decreases, although the adsorption on the charged surfaces is slightly higher than in the neutral walls case, at $pH_2$ and $pH_3$. As expected, the adsorption of counterions on the negatively charged wall increases with their concentration. As in Figure 3, the polyelectrolytes and the counterions are evidently in competition with each other for the adsorption sites on the surfaces. A comparison with experimental results is difficult because most works report experiments performed under conditions different from those of the simulations. However, our simulations can be compared with the adsorption of negative polyelectrolytes on a positively charged surface, i.e., with polyelectrolytes adsorbed onto oppositely charged surfaces, as in Figure 4. One can compare the isotherms in Figure 4 with the adsorption of weak anionic polyelectrolytes on a positively charged surface whose charge is independent of pH [11], as is the case in our present work. In reference [11] the experimental adsorption was found to decrease while the degree of ionization of anionic polyelectrolytes increased. Those adsorption isotherms follow the same trends as the results presented in Figure 4 since the predicted polyelectrolyte adsorption decreases also with the increasing in the degree of ionization. Thus, the adsorption isotherms shown in Figure 4 reproduce very well the experimental results [11].

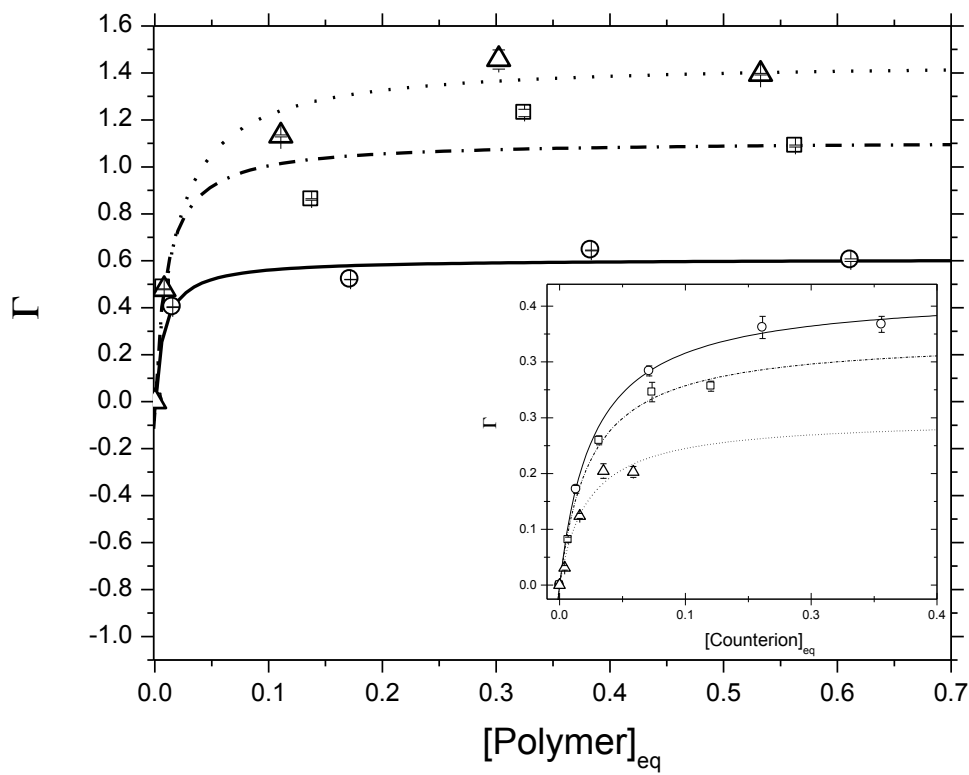

**Figure 5.** Adsorption isotherms of anionic polyelectrolytes on negatively charged surfaces for $\alpha_1$ ($\Delta$), $\alpha_2$ ($\square$) and $\alpha_3$ ($\circ$). In this case $pH_1 < pH_2 < pH_3$, and as before, $\alpha_1 < \alpha_2 < \alpha_3$. The adsorption isotherms of the positive counterions are shown in the inset. The statistical uncertainty is smaller than the symbols size. The continuous lines represent the data fitting to Langmuir's model, Equation (24).

Let us now consider the case when walls and polyelectrolytes have the same kind of electrostatic charge. The adsorption isotherms of the negative polyelectrolytes (and the positive counterions) on negatively charged surfaces are shown in Figure 5, and these isotherms follow the *same* trends as those of the 2 preceding figures when they are interpreted in terms of the degree of ionization. That is because the relation between the degree of ionization and pH is given by the positive sign in equation (25), namely, $pH_1 < pH_2 < pH_3$ when $\alpha_1 < \alpha_2 < \alpha_3$. Therefore, as Figure 5 shows, anionic polyelectrolyte adsorption decreases as the pH is increased, while the counterions adsorption increases with the pH. Hence, the adsorption isotherms in Figure 5 represent the same behavior as that seen in Figures 3 and 4, when the degree de ionization of polyelectrolytes is considered, namely the greater the ionization, the lower the polyelectrolyte adsorption, independently of the nature of the polyelectrolyte.

As the pH is increased the degree of ionization of the negative polyelectrolyte is greater, and the presence of positive counterions rises as well. Thus, the adsorption of the negative polyelectrolyte on the negative wall is reduced as the pH is increased, while the adsorption of the positive ions is favored. However, the latter is not enough to induce a surface charge inversion that may favor the adsorption of the likely charged polyelectrolyte. The results shown in Figure 5 are then what would be expected for these systems. Others have pointed out also the relevance of the counterions in the adsorption phenomenon. For example, the role of the ions and counterions has been evidenced in the adsorption of spherical macro ions on charged walls [43]. These authors have shown that the combination of steric and electrostatic effects of ions, counterions and macro ions may produce adsorption on neutral, or on oppositely and a likely charged surfaces, as seen in our simulations. The density profiles of the monomers that make up the polyelectrolytes for the systems studied in Figure 3 – 5 confirm the arguments given above, as seen in Figure 6.

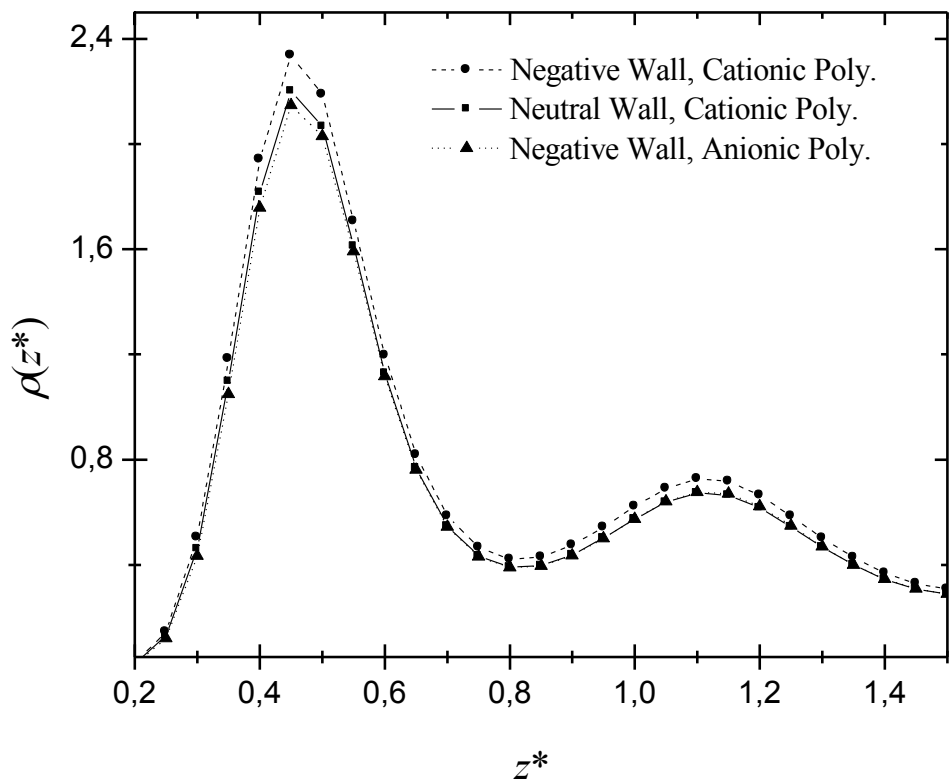

**Figure 6**. Segment density profiles for cationic polyelectrolytes on neutral walls (solid line and squares), cationic polyelectrolytes on negatively charged walls (dashed line, solid circles), and anionic polyelectrolytes on negatively charged walls (dotted line, solid triangles), all at pH$_3$. The polyelectrolyte concentration in each case is 1.4 monomers/nm$^3$. Only the structure of the density profiles close to the left wall is shown, and the solvent and counterions density profiles are omitted, for clarity. The density profiles on the right wall are completely symmetric with respect to those of the left one therefore they are omitted also.

The density profiles in Figure 6 show that although the adsorption on the surfaces are quite similar for the systems studied in Figures 3 – 5, some differences remain. The most important feature is the height of the first maximum in Figure 6, which corresponds to the density of the first layer of adsorbed polyelectrolytes. The largest adsorption occurs when the system is composed of positively charged (cationic) polyelectrolytes confined by negatively charged surfaces (solid circles, dotted line in Figure 6). It is followed by the adsorption of cationic polyelectrolytes on neutral walls (solid squares, solid line), and the smallest adsorption maximum corresponds to the negatively charged polyelectrolytes on negatively charged surfaces (dashed line in Figure 6). The trend shown by the height of the first maximum in the density profiles shown in Figure 6 indicates that there are a few more monomers of the cationic polyelectrolyte on the negatively charged wall because of their electrostatic attraction, while there are less of the anionic one precisely because of their Coulomb repulsion for the wall. However, it appears that electrostatic interactions cannot be solely responsible for the adsorption isotherms shown in the previous figures, since even when the surfaces are electrically neutral one sees a considerable adsorption of (cationic) polyelectrolytes in Figure 6. We stress here that even in the absence of electrostatic interactions, if the neutral wall – polymer parameter, $a_w$ (see equation (10)), is less repulsive than it is between the wall and the solvent, then polymer adsorption will be favored.

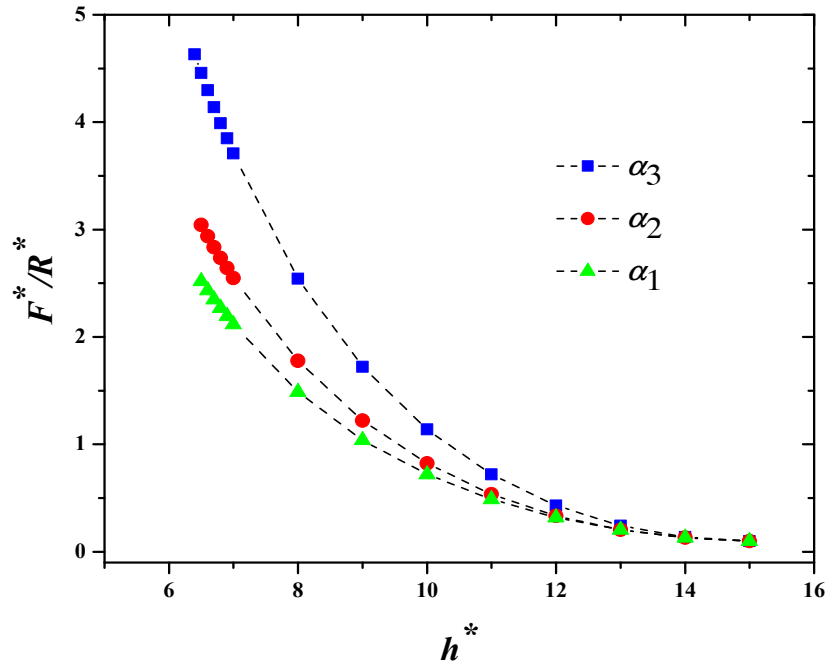

**Figure 7**. Full surface forces ($F^*$) between two negatively charged walls of equal curvature radius ($R^*$) with cationic polyelectrolytes adsorbed on the surfaces, for the 3 different ionization degrees ($\alpha$) used also in Figure 4, where $\alpha_1 < \alpha_2 < \alpha_3$ with $pH_1 > pH_2 > pH_3$, at a fixed polyelectrolyte concentration of 20 molecules per volume, as a function of distance between surfaces ($h^*$). Asterisks indicate adimensional magnitudes. The lines are only guides for the eye.

Many-body surface forces are also of major importance when studying confined complex fluids, as they can be used to ascertain the stability of the dispersion, and can be measured, for example, with AFM [2]. Therefore this is a property that leads directly to predictions, which can be tested experimentally. Figure 7 shows the surface force (per curvature radius, $R$, of the AFM probe) isotherms obtained as the ionization degree (and consequently, pH) is varied, for cationic polyelectrolytes adsorbed on negatively charged surfaces. There are 20 polyelectrolytes for each wall-to-wall separation; the same pH values used for the adsorption isotherms shown in Figures 3 and 4 were used for the prediction of surface force isotherms. The trend observed is that the force becomes more repulsive, the larger the ionization degree, $\alpha$. Note that in this case, as $\alpha$ increases the pH is reduced, and $pH_1 > pH_2 > pH_3$. These are the first surface force isotherms of this kind (using $\alpha$ as a variable) predicted by simulation, to our knowledge. For the calculation of the surface force we started out by calculating the normal pressure tensor ($P_{zz}$), using the virial theorem [32], at a given maximum surface-to-surface separation, $h$. Afterward, the distance $h$ was reduced and $P_{zz}$ was recalculated; this process was repeated for as many simulation box volumes as possible. Then, using the Derjaguin approximation [2], the full surface force per curvature radius ($F/R$) can be obtained, as follows:

$$\frac{F}{R} = -\pi \int \left[ P_{zz}(z) - P_b \right] dz. \qquad (26)$$

In equation (26), $P_b$ is the pressure of the bulk, i.e., unconfined fluid. We have emphasized this is a full, many-body force because it contains the contributions from solvation, entropic and electrostatic forces. One of the most salient features of the curves in Figure 7 is that all forces are repulsive; moreover, the larger the ionization degree, the stronger the repulsive force. This trend indicates that the most thermodynamically stable colloidal dispersion is the one with the most dissociated polyelectrolytes, i.e., those with the lowest pH values for cationic polyelectrolytes, as in Figure 7. At those values the electrostatic interaction is rather strong, which allows the adsorbed polyelectrolyte layer to swell, giving rise to a larger repulsive force. By contrast, at high pH values the cationic layer is collapsed as the interaction is mostly of steric origin, allowing the colloidal particles to come closer together, that is, lowering the surface force. Notice also how the difference between the surface forces of colloidal dispersions with polyelectrolytes at $\alpha_2$ and $\alpha_1$ is smaller than that between the dispersions at $\alpha_3$ and $\alpha_2$. That is due to the fact that the ionization degree difference between $\alpha_3$ and $\alpha_2$ is larger than that between $\alpha_2$ and $\alpha_1$. The largest repulsion occurs at the smallest distances separating any 2 colloidal particles coated with polyelectrolytes because at those distances the polyelectrolyte brushes can overlap. The trends shown in Figure 7 indicate the regimes where the surface force is due mainly to electrostatic repulsion (e. g., at $\alpha_3$), competition/cooperation between electrostatic and steric interactions (say, at $\alpha_2$), or mainly short-range, steric repulsion (at $\alpha_1$). At large interparticle distances the latter are zero because of their short-range nature, while the Coulomb interactions are negligible due to screening through the counterions; therefore the surface forces become negligible at those distances, as expected and observed in experiments [2].

Recent AFM measurements [44] on silica particles and surfaces coated with weak anionic polyelectrolyte brushes made of poly(acrylic acid), in KCl solutions at various pH values have shown precisely the trends we have predicted in Figure 7. This conclusion follows because the range and the strength of the repulsive force increase with increasing $\alpha$, given that the anionic polyelectrolyte dissociation increases with the pH. At acid pH, the measured surface force isotherms [44] exhibited strong steric repulsion up to distances of the order of the polyelectrolyte brush length; while at basic pH values it was argued that the polyelectrolyte brushes were well dissociated giving rise to purely electrostatic repulsion. Claesson and Ninham [45] carried out surface force measurements as a function of pH using a surface force apparatus, between chitosan (a cationic polyelectrolyte) coated mica (negatively charged) surfaces, finding exactly the same trends (even quantitatively, when $F^*/R^*$ and $h^*$ are properly dimensionalized) we predict in Figure 7 for a qualitatively similar system. These authors [45] found that their results could be explained assuming the chitosan layers adopt a flat conformation on the mica surfaces, which is precisely how our model polyelectrolytes adsorb. In this sense it should be stressed that our polyelectrolytes are "surface modifying" polyelectrolytes, rather than brushes, but the interplay of the segment-segment and segment-surface interactions is the same for both types [46, 47].

Also, because our model surface – modifying polyelectrolytes tend to be completely adsorbed on the particle's surface (at the concentration used in the calculation of the force isotherms), there are no remaining dangling segments in the solvent, which could then associate with other dangling segments on the opposite surface, thereby forming a "bridge". That is the reason why we do not see an attractive bridging effect [48] in our force profiles, and that is also why the adsorption isotherms can be fit relatively well to the monolayer Langmuir model [39].

At this point it is instructive to recall that polyelectrolyte brushes are usually added to colloidal particles to impart them with appropriate characteristics so that the particle dispersion will remain stable. It is then of notice that although the largest adsorption was obtained for polyelectrolytes at the smallest ionization degree (or largest pH, which is $pH_1$, solid triangles in Figure 4), such adsorption does not necessarily translate into the most stable dispersion, as observed in Figure 7, where the lowest surface force curve is obtained for the largest pH ($\alpha_1$). It may be possible to obtain larger surface forces, i.e., better colloidal stability, with thinner polyelectrolyte layers adsorbed on the colloidal particles because the repulsive interactions (be they of steric or of electrostatic origin) act more effectively if the appropriate electrochemical environment (namely, the pH) is provided.

## VIII. CONCLUSIONS

Adsorption and surface force isotherms of polyelectrolytes adsorbed on charged and neutral surfaces were obtained using a mesoscopic, short-range interaction model in MC simulations at fixed chemical potential, volume and temperature. The results showed that the ionization degree of the polyelectrolyte, and therefore its pH, is a key factor when it comes to determining its adsorption characteristics. We predict the same trends in the adsorption isotherms for anionic and for cationic polyelectrolytes in athermal solvent, when adsorbed on negatively charged surfaces as function of their degree of ionization. The main trend is that the lower the polyelectrolyte ionization degree, the greater the adsorption, independently of the nature of the polyelectrolyte (positive or negative). Our results indicate that a general trend in the adsorption of polyelectrolytes on charged or neutral surfaces exists, which is supported by experimental trends taken from the recent literature. The role of the counterions in the process of adsorption is also evident and follows an opposite behavior to that of the polyelectrolytes, namely that at high (low) polyelectrolyte adsorption, the counterion adsorption is low (high). This indicates there is competition for the adsorption sites, a phenomenon that has scarcely been considered. Additional, we have shown that the largest adsorption of a polyelectrolyte does not necessarily translate into the best stability of a colloidal suspension as obtained from surface force profiles, and argue that this is due to the range of the interactions. Electrostatic interactions do not appear to be the sole driving force in the process of adsorption, but rather there exists a complex competition of screening and steric effects. On the other hand, electrostatics plays a major role in increasing the surface force, when the ionization degree is increased. DPD has proved to be useful when complemented with appropriately adapted Coulomb interactions not only among polymers, but on the confining surfaces also. This work should be of interest to specialists in stimuli-responsive materials and workers in the soft matter community, among others.

ACKNOWLEDGEMENTS

This work was sponsored by the Centro de Investigación en Polímeros (Grupo COMEX). The authors are grateful to M. A. Balderas Altamirano for useful discussions, to IPICyT for granting us use of the computational facilities, and to *Red Temática Materia Condensada Blanda* – CONACyT.